\documentclass[sigconf,authorversion]{acmart}

\usepackage[utf8]{inputenc}
\usepackage{xcolor}
\usepackage{hyperref}
\hypersetup{hidelinks}
\usepackage{booktabs}
\usepackage{pifont}%
\usepackage{balance}
\newcommand{\cmark}{\ding{51}}%
\newcommand{\xmark}{\ding{55}}%

\newcommand{\cn}{\textsc{MultiMediate\,}}

\AtBeginDocument{%
  \providecommand\BibTeX{{%
    \normalfont B\kern-0.5em{\scshape i\kern-0.25em b}\kern-0.8em\TeX}}}

\copyrightyear{2022} 
\acmYear{2022} 
\setcopyright{acmlicensed}\acmConference[MM '22]{Proceedings of the 30th ACM International Conference on Multimedia}{October 10--14, 2022}{Lisboa, Portugal}
\acmBooktitle{Proceedings of the 30th ACM International Conference on Multimedia (MM '22), October 10--14, 2022, Lisboa, Portugal}
\acmPrice{15.00}
\acmDOI{10.1145/3503161.3551589}
\acmISBN{978-1-4503-9203-7/22/10}

\begin{document}

\author{Philipp M\"uller}\thanks{$^*$These authors contributed equally to this work}
\affiliation{
    \institution{German Research Center for Artificial Intelligence}
    \city{Saarbr\"ucken}
    \country{Germany}
}
\email{philipp.mueller@dfki.de}
\author{Michael Dietz$^{*}$}%
\affiliation{
    \institution{University of Augsburg}
    \city{Augsburg}
    \country{Germany}
}
\email{michael.dietz@informatik.uni-augsburg.de}
\author{Dominik Schiller$^{*}$}%
\affiliation{
    \institution{University of Augsburg}
    \city{Augsburg}
    \country{Germany}
}
\email{dominik.schiller@informatik.uni-augsburg.de}

\author{Dominike Thomas$^{*}$}%
\affiliation{
    \institution{University of Stuttgart}
    \city{Stuttgart}
    \country{Germany}
}
\email{dominike.thomas@vis.uni-stuttgart.de}
\author{Hali Lindsay}%
\affiliation{
    \institution{German Research Center for Artificial Intelligence}
    \city{Saarbr\"ucken}
    \country{Germany}
}
\email{hali.lindsay@dfki.de}
\author{Patrick Gebhard}
\affiliation{
    \institution{German Research Center for Artificial Intelligence}
    \city{Saarbr\"ucken}
    \country{Germany}
}
\email{patrick.gebhard@dfki.de}
\author{Elisabeth Andr\'e}
\affiliation{
    \institution{University of Augsburg}
    \city{Augsburg}
    \country{Germany}
}
\email{andre@informatik.uni-augsburg.de}
\author{Andreas Bulling}
\affiliation{
    \institution{University of Stuttgart}
    \city{Stuttgart}
    \country{Germany}
}
\email{andreas.bulling@vis.uni-stuttgart.de}

\title{\cn'22: Backchannel Detection and Agreement Estimation in Group Interactions}
\date{June 18, 2022}

\renewcommand{\shortauthors}{Philipp Müller et al.}

\begin{abstract}

Backchannels, i.e. short interjections of the listener, serve important meta-conversational purposes like signifying attention or indicating agreement.
Despite their key role, automatic analysis of backchannels in group interactions has been largely neglected so far.
The \cn challenge addresses, for the first time, the tasks of \textit{backchannel detection} and \textit{agreement estimation from backchannels} in group conversations.
This paper describes the \cn challenge and presents a novel set of annotations consisting of 7234 backchannel instances for the MPIIGroupInteraction dataset.
Each backchannel was additionally annotated with the extent by which it expresses agreement towards the current speaker.
In addition to a an analysis of the collected annotations, we present baseline results for both challenge tasks.

\end{abstract}

\ccsdesc[500]{Computing methodologies~Artificial intelligence}

\keywords{challenge, dataset, backchannel detection, agreement estimation}

\settopmatter{printfolios=true}

\maketitle

\section{Introduction}

Backchannels, i.e. short interjections made by listeners,
are at the core of the bilateral nature of dialogue~\cite{clark_speaking_2004}.
That is, dialogue in which a listener's responses affect the speaker's speech acts.
They may consist of a verbal phrase like ``oh yes'', a sound (e.g. ``hm''), or bodily gestures, such as head nods or hand movements, or combinations thereof.
Backchanneling serves the important functions of confirming listeners' attention and comprehension~\cite{gravano_backchannel-inviting_2009}, as well as regulating turn-taking~\cite{schegloff_discourse_1982}.
In addition, it is used to communicate agreement or disagreement with the current speaker and is therefore an important indicator of participants' opinions and intra-group relations~\cite{allbeck_multimodal_2010}.
The inability to appropriately perform backchanneling can have severe impact on dialogue, e.g. by distracting the speaker~\cite{park_backchannel_2017, bavelas_listeners_2000}.

As a result, automatic analysis of backchanneling behaviour is crucial and has significant potential for artificial systems designed to passively monitor or actively mediate human dialogue~\cite{schiavo2014overt,balaam2011enhancing,park2020investigating,penzkofer2021conan}.
For example, artificial mediators~\cite{birmingham2020can,engwall2020interaction,bohus2010facilitating} could analyse the frequency of backchannels to gauge participants' engagement and encourage those who are disengaged from the discussion.
Agreement or disagreement expressed in backchannels could also help artificial mediators to better understand the opinions of participants.
This may enable mediators to give a voice to participants who find it difficult to express diverging opinions.
Despite this potential, automatic analysis of backchannels in group interactions remains largely unexplored due to the lack of suitable datasets.

With \cn'22, we present the first challenge for automatic backchannel analysis in group interactions.
To this end, we introduce the first publicly available dataset of backchanneling behaviour in group discussions.
We fully annotated the MPIIGroupInteraction dataset~\cite{muller_detecting_2018} with 7234 backchannel instances.
Furthermore, each backchannel instance was rated with the extent to which it expresses agreement with the current speaker.
We present analyses of our novel annotations as well as evaluations of baseline approaches for \cn'22.
All collected annotations, baseline implementations, and raw feature representations are made publicly available for further use.\footnote{\url{https://multimediate-challenge.org}}

\section{Previous Work}

\subsection{Backchannel and Agreement Datasets}

While a number of different datasets were recorded for the purpose of backchannel analysis~\cite{cafaro:et:al:2017,prevot_cup_2016, bavelas_listeners_2000,morency_predicting_2008}, they are rarely public and not sufficient to ensure progress on automatic backchannel analysis in group interactions.
In Table~\ref{tab:datasets} we list recent publicly available datasets that were used in BC research, the majority of which consists of spoken dialogues.
Note that while these datasets are publicly available, the backchannel annotations collected on them are often not (column ``Pub.'' in Table~\ref{tab:datasets}).
Furthermore, all existing datasets (except the Canal9 Corpus~\cite{vinciarelli2009canal9,poggi_types_2010} for which no publicly available backchannel annotations exist) consist of dyadic interactions and are thus not fitting for group behaviour analysis. 

In terms of annotations of backchannels, some studies use semi-automatic labeling of potential events~\cite{baur:et:al:2020,boudin_multimodal_2021}, or query a dataset for specific BC keywords~\cite{ortega_oh_2020}. 
Backchannel annotations are most often based on a set list of backchannel events (``is there a ``uh-huh" here?")~\cite{ortega_oh_2020, boudin_multimodal_2021, skantze_turn-taking_2014, ferre_unimodal_2017, mueller_using_2015},
and only rarely on the holistic perception of a backchannel (``is there a backchannel here?")~\cite{heldner_backchannel_nodate, allwood_study_2003, bavelas_listener_2002}. %

While a relationship between backchannels and agreement has been discussed~\cite{schegloff_discourse_1982}, few studies have systematically investigated it. ~\cite{prevot_cup_2016} included agreement in their annotations of backchannels, but did not analyse it; ~\cite{galley_identifying_2004, hillard_detection_2003} distinguish backchannels from agreement signals.
Some studies of agreement include backchannels~\cite{bousmalis_spotting_2009, bousmalis_towards_2013}, but to the best of our knowledge, no study of backchanneling investigated agreement transmitted \textit{through} the backchannels themselves.

In our work, we present the first publicly available annotations of backchannel occurances and agreements expressed via backchannels in group interactions.
Our full dataset is manually annotated and with over 33 hours of annotated human behaviour across training, validation and test sets, its size is equal to the largest dyadic BC dataset currently available (see Table \ref{tab:datasets}).

\begin{table}[t]
\begin{tabular}{llllllll}
\toprule
 \textbf{Name} &  \textbf{Pub.} & \textbf{Part.}  & \textbf{Size} & \textbf{Lang.} \\
 \midrule
Cheese-Paco~\cite{blache_integrated_2020, boudin_multimodal_2021}  &\xmark & 2  & 2h & FR \\
Vyaktitv~\cite{jain_exploring_2021}  &\xmark &2 & 14h & HI \\
Spontal~\cite{heldner_backchannel_nodate}   &\xmark & 2 &0h40 & SV \\
P2PSTORY~\cite{singh_p2pstory_2018}   &\cmark &2 &2h30 &EN \\ 
Canal9~\cite{poggi_types_2010}  &\xmark &5 &-- &FR \\
Cup of CoFee~\cite{prevot_cup_2016}  &\cmark &2 &33h42 &FR \\
IFADV~\cite{truong_multimodal_2011}  &\xmark &2 &9h30 &NL \\
Spoken Language~\cite{allwood_study_2003, allwood_swedish_nodate}   &\xmark &2 &-- &SV \\
NOXI~\cite{cafaro:et:al:2017}  & \xmark &2 &25h18 & 7 Lgs\\
\midrule
MPIIGroupInteraction~\cite{muller_detecting_2018}  &\cmark &3-4 &33h40 &DE \\
\bottomrule
\end{tabular}
\caption{Publicly available audio-visual human-human interaction datasets used in BC research. 
\textit{Pub.} indicates whether BC annotations are publicly available;
\textit{Part.} the number of participants per interaction. 
\textit{Size} is the duration of individual human behaviour annotated with BCs (where reported). E.g. for a dyadic dataset it is twice the length of annotated interactions. \textit{Lang.} indicates the language of the dataset.
}
\label{tab:datasets}
\end{table}

\subsection{Computational Models}

The prediction (i.e. anticipation) of backchannels is a highly active area of research in social signal processing.
One major motivation is the goal to generate natural backchanneling behaviour in artificial agents.
There is a variety of traditional machine learning methods used for predicting backchannels~\cite{morency_predicting_2008,fujie_back-channel_2005, goswami_towards_2020, ferre_unimodal_2017,kawahara_prediction_2016,terrell_regression-based_2012}.
Recently, 
LSTM networks became the most frequent choice~\cite{ruede_yeah_2019, hara_prediction_2018, adiba_towards_2021, dharo_delay_2021}, along with residual networks~\cite{goswami_towards_2020}. 
Multitask learning also seems to be a particularly successful approach~\cite{ishii_multimodal_2021,jang_bpm_mt_2021,hara_prediction_2018}.
The most common features used in BC prediction are prosodic~\cite{ruede_enhancing_2017, ruede_yeah_2019, hara_prediction_2018, adiba_towards_2021} while some research also makes use of linguistic features such as word embeddings~\cite{ruede_enhancing_2017, ruede_yeah_2019, adiba_towards_2021, dharo_delay_2021} or syntactic (part of speech tags), semantic (concreteness, valence) and discourse features \cite{blache_integrated_2020, kawahara_prediction_2016, boudin_multimodal_2021}. Research shows that adding lexical features to accoustic ones improves results~\cite{ortega_oh_2020, ruede_enhancing_2017}. \cite{gravano_backchannel-inviting_2009} offers a review of the most important backchannel-inviting cues which may also be used as features. 

In contrast to backchannel prediction, we define the task of backchannel detection as categorizing observed behaviour as to whether a backchannel was shown or not.
While to the best of our knowledge, this task was not studied in isolation in previous work,
backchannel detection can appear as a part of the multi-class problem of dialogue act classification (DAC)~\cite{saha2020towards,alexandersson1997dialogue}.
DAC is primarily addressed by text analysis~\cite{grau2004dialogue,khanpour2016dialogue,raheja2019dialogue} and only few works incorporated multi-modal information like emotional expressions~\cite{saha2020towards,boyer2011affect}.
Importantly, no previous work addressed the task of backchannel detection in group interactions from multi-modal behaviour.

With \cn'22 we aim to attract researchers to two challenging problems centered around backchannels in group discussions: backchannel detection and estimation of the amount of agreement expressed in a backchannel.
In line with human studies that underline the importance of non-verbal backchannel cues~\cite{allwood_study_2003,kendon_functions_1967}, we provide annotations of backchannel behaviour that can take place both via speech and in the visual domain.
We contribute to improved comparability between approaches by using an unpublished test set for evaluation.

\section{Challenge Description}
As in \cn'21~\cite{muller2021multimediate}, our challenge is based on the MPIIGroupInteraction dataset~\cite{muller_detecting_2018,muller2018robust}.
This dataset has been used for diverse tasks, including low rapport detection~\cite{muller_detecting_2018}, emergent leadership detection~\cite{muller2019emergent}, eye contact detection~\cite{muller2018robust,fu2021using}, next speaker prediction~\cite{birmingham2021group}, and body language detection~\cite{balazia2022bodily}.
For \cn'22 we collected novel backchannel annotations on the whole dataset.
Test samples (excluding ground truth) are released to participants before the challenge deadline.
Participants in turn submit their predictions for evaluation by the challenge organisers.
We first describe MPIIGroupInteraction and subsequently discuss annotation procedures and task definitions for backchannel detection and agreement estimation.

\subsection{Dataset}

\begin{figure}[t]
  \centering
  \includegraphics[width=\columnwidth]{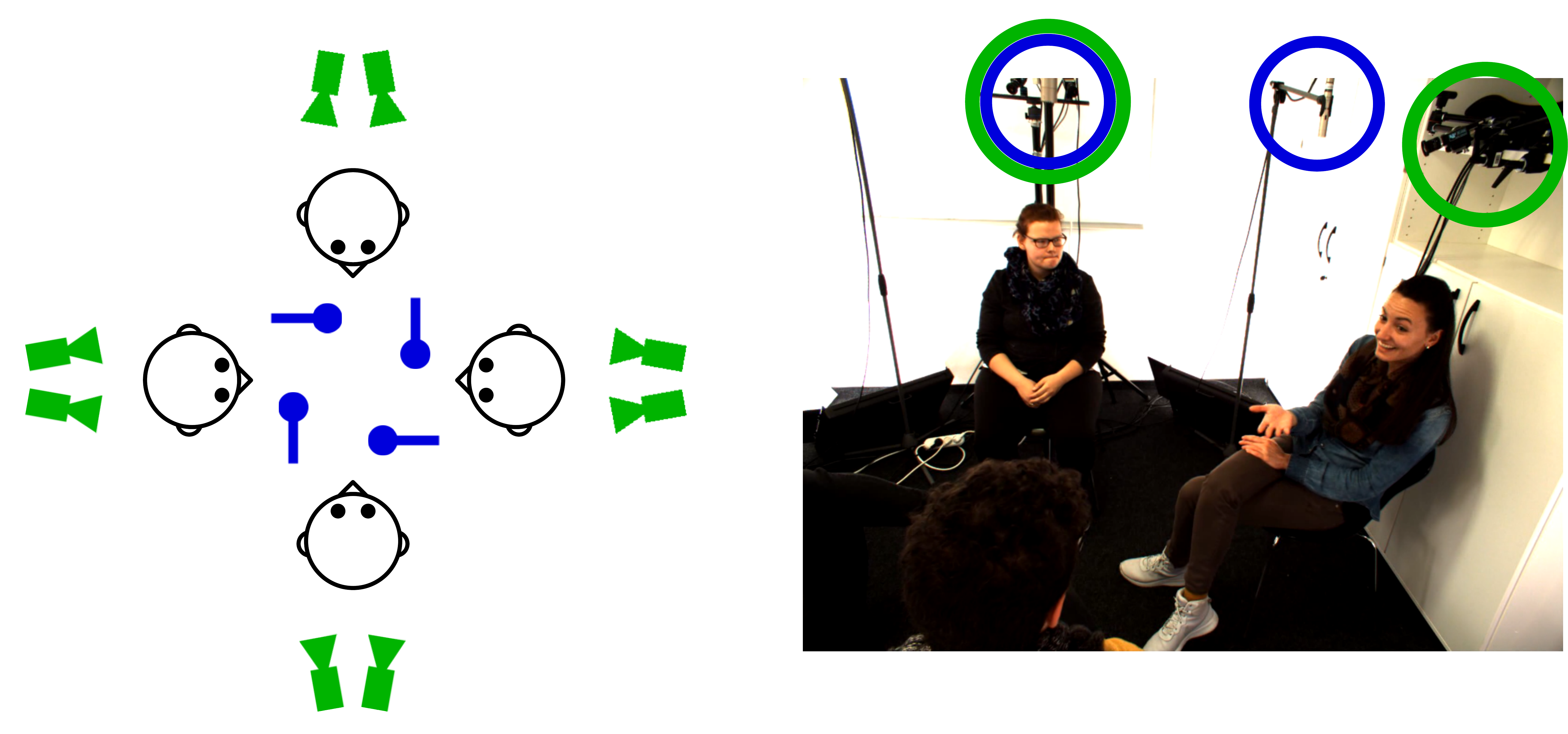}
  \caption{The recording setup for MPIIGroupInteraction. %
  Printed with permission from the authors of~\cite{muller_detecting_2018}.
  }~\label{fig:study_design_iui18}
\end{figure}

\begin{figure*}[t]
    \centering
    \includegraphics[width=2\columnwidth ]{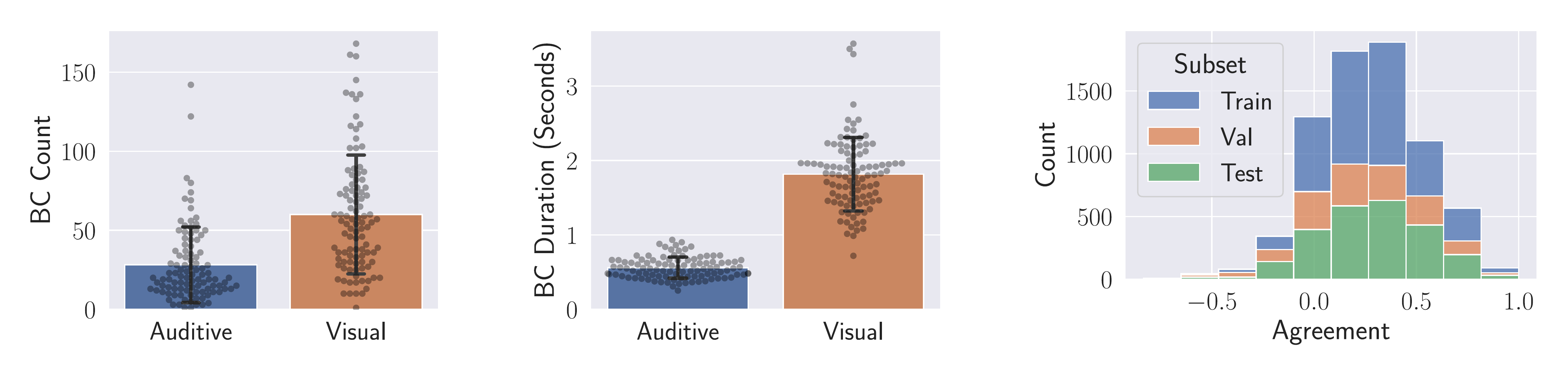}
    \caption{\textbf{Left:} Number of auditive and visual backchannel annotations per participant. Each point represents one participant, bars and whiskers represent the mean and standard deviation across participants. \textbf{Middle:} Average duration of auditive and visual backchannels. \textbf{Right:} Distribution of agreement scores across \cn'22 train, val, and test samples.}
    \label{fig:annotations}
\end{figure*}

\paragraph{Training data.}
We use the publicly available recordings of MPIIGroupInteraction~\cite{muller_detecting_2018} as training data.
The dataset comprises 22 conversations between three to four people on a maximally controversial topic, lasting 20 minutes each.
While discussing the topic, interactants were recorded with four microphones and eight frame-synchronised video cameras (see Figure~\ref{fig:study_design_iui18}).

\paragraph{Evaluation data.}
\label{sec:eval_data}
We follow \cn'21~\cite{muller2021multimediate} and use six yet unpublished discussions that were recorded during the creation of MPIIGroupInteraction~\cite{muller_detecting_2018} for testing.
These six discussions followed the same procedure as the training discussions with the only exception that the topic was not chosen to be maximally controversial.
Instead, a topic was randomly selected for each group and participants were asked to take on opposing views for themselves.

\subsection{Backchannel Detection Task}
\label{bcdetection}
\paragraph{Backchannel Annotations}
In a first step annotators where asked to label the occurrences of backchanneling behaviour with respect to different modalities: \textit{auditive}, and \textit{visual}. 
Visual annotations take only the aspects observable in the video into account (e.g. nodding or head shaking).
Auditive backchanneling behaviour, in turn, relies only on cues from the audio signal, encompassing verbal (e.g. "yes", "oh really?") and paraverbal behaviour (e.g. "uhm hum", "aha").
Each of three annotators labeled a specific portion of the dataset with respect to backchannels for each modality. 
\autoref{fig:annotations} (left) shows the average of annotated backchannel events annotated per participant in each modality.
On average, we observed 28 annotations per participant for the auditive modality, and 60 for the visual modality.
\autoref{fig:annotations} (middle) shows the average duration of annotated backchannel events for each modality.
Visual backchannel events are on average longer (1.8 seconds) compared to auditive backchannel events (0.56 seconds).
The modality specific labels were subsequently joined using the logical \textit{OR} operator. 
Hence, overlapping backchannels in different modalities (e.g. a person nods while simultaneously saying "yes") are merged to a new, modality independent \textit{backchannel} label. 
To provide negative examples we also calculated an equal amount of non-backchannel samples per session. 
Overall this results in 14468 labels partitioned into the following splits: 6716 Train, 2854 Val, 4898 Test.

\paragraph{Task definition}
Given an observation window of 10 seconds participants have to detect if there is a backchannel present in the sample. 
Every sample has been created such that the end of an annotated backchannel has been used as the end the sample or that a sample consists of 10 seconds with no annotated backchannels. 
The performance metric for the task is accuracy.

\subsection{Backchannel Agreement Estimation Task}
\paragraph{Agreement Annotations}

In the second step, all backchannel instances that where annotated for the \textit{backchannel detection test} were labeled with respect to their level of expressed agreement on a scale from -1 (total disagreement) to 1 (total agreement) using a step size of 0.1.  
In case annotators found that the instance was erroneously labeled as a backchannel, they indicated this fact with an extra label. 
When all annotators agreed that an instance was wrongly labeled as a backchannel the respective sample has been removed from the dataset during the sampling process.
This annotation step has been performed by each annotator for all backchannel labels (see section \ref{bcdetection})
The groundtruth labels were then created by averaging all three annotations per sample. 
Overall this results in 7234 labels partitioned into the following splits: 3358 Train, 1427 Val, 2449 Test.
To quantify the reliability of agreement annotations, we compute the average Spearman $\rho$ when comparing a left-out annotator to the average of the two remaining ones is.
We observe a $\rho$ of 0.62, indicating substantial agreement.
\autoref{fig:annotations} (right) shows the distribution of aggregated agreement annotations.
The distribution is centered in the positive range between 0 and 0.5.
The detection of disagreement represents a special challenge due to the small number of samples below 0.

\paragraph{Task definition}
For the backchannel agreement estimation tasks participants have to predict the average expressed agreement per sample on scale from -1 to 1. 
The performance metric for the task is mean squared error.

\section{Experiments and Results}

\subsection{Features}
We extracted the same set of features for backchannel detection and agreement estimation from backchannels.
Features were extracted from the last second of the 10 second input window and aggregated by computing the mean or the mean of absolute differences (``mean delta'') of adjacent frames over this second.

\subsubsection{Head Features} We extracted features from participants' head and face using OpenFace 2.0~\cite{baltrusaitis2018openface}. These include mean and mean delta of AU intensity estimates, mean delta of head orientation and translation, as well as mean delta of gaze angles for both eyes.
In total, we extracted 46 features based on OpenFace 2.0.

\subsubsection{Pose Features} We extract body pose estimates using OpenPose~\cite{cao2017realtime} and employ a set of angular features that proved successful for group interaction analysis~\cite{beyan2017moving}.
These features consist of angles between body parts, e.g. the angle defined by the line between left shoulder and left elbow and the line between left shoulder and right shoulder.
We limit ourselves to the ``Upper Body'' and ``Head'' features described in \cite{beyan2017moving}, as lower body pose estimates tend to be unreliable on the MPIIGroupInteraction dataset.
We compute the mean delta of these features, resulting in 8 body pose features.

\subsubsection{Voice Features} We extracted the extended Geneva Minimalistic Acoustic Parameter Set (eGeMAPS)~\cite{eyben2015geneva} on the last second of the input window.
This set consists of 88 acoustic parameters that are commonly applied to tasks like depression, mood, and emotion recognition~\cite{valstar2016avec}, or Alzheimer's Dementia recognition~\cite{luz2020alzheimer}.

\subsection{Prediction Approach}
For backchannel detection, we trained a binary Support Vector Classifier (SVC) with rbf kernel.
For the agreement estimation task, we trained a Support Vector Regressor (SVR) with rbf kernel.
We employed 10-fold cross-validation on the training set to choose $\gamma$ and $C$ parameters of the SVC/SVR.
For test set evaluations, we trained on training and validation sets; for evaluations on the validation set we only trained on the \cn'22 training set.

\subsection{Results}
We present evaluation results for different combinations of feature sets in \autoref{tab:results}.
For fairness reasons, we only evaluate two feature sets for each task on the training set: the best performing feature set on the validation set, as well as all included features.

\subsubsection{Backchannel Detection}
The best feature set on the validation set consisted of a combination of head and pose features, reaching 0.639 accuracy on validation and 0.596 accuracy on the test set.
This clearly outperformed the trivial baseline of a random predictor at 0.5 accuracy and was marginally better than all featuresets combined.
Notably, our experiments on the validation set revealed that each individual feature set achieved above-random performance on backchannel detection, even though OpenFace 2.0 based head features were clearly leading.
An ablation of the head feature set revealed that head pose alone (i.e. mean delta of translation and rotation of the head) reaches 0.636 accuracy.
This is likely due to the strong association between backchanneling and nodding.

\subsubsection{Agreement Estimation}
For agreement estimation, head pose features performed best on the validation set (0.075 MSE) and reached 0.061 MSE on the test set.
In both cases this improves above the trivial baseline of using the mean on the training set as a predictor.
However, featuresets that do not include head pose or gaze features fail to outperform the trivial baseline.
This is in contrast to the backchannel detection task and indicates the difficulty of backchannel agreement estimation.

\begin{table}[t]
\small
\centering
    {\begin{tabular}{l  c  c  c c}
        \toprule
         & Detection & Detection & Agreement & Agreement \\ 
        Features  & Val ACC~$\uparrow$ & Test ACC~$\uparrow$ & Val MSE~$\downarrow$ & Test MSE~$\downarrow$ \\
        \midrule
        Head & 0.621 & -  & 0.079 & - \\
        \ \ AUs only & 0.591 & - & 0.085 & -\\
        \ \ H. Pose only & 0.636 & - & \textbf{0.075} & \textbf{0.061}\\
        \ \ Gaze only & 0.622 & - & 0.078 & -\\
        Pose & 0.531 & - & 0.086 & - \\
        Voice & 0.567 & - & 0.085 & - \\
        Head + Pose & \textbf{0.639} & \textbf{0.596} & 0.079 & - \\
        All Features & 0.636 & 0.592 & 0.079 & 0.064 \\
        \midrule
        Trivial Basel. & 0.500 & 0.500 & 0.085 & 0.066 \\
        \bottomrule
    \end{tabular}}
    \caption{Validation and test results for backchannel detection and agreement estimation from backchannels.} 
    \label{tab:results}
\end{table}

\section{Conclusion}
We introduced \cn'22, the first challenge addressing backchannel detection and agreement estimation from backchannels in well-defined conditions and evaluated baseline approaches for each task.
In addition we introduced a novel publicly available dataset of backchannel annotations in group interactions that is a valuable resource for research on backchannel detection and agreement estimation, even beyond the \cn challenge.

\begin{acks}
P. M\"uller and H. Lindsay were funded by the \grantsponsor{01IS20075}{German Ministry for Education and Research (BMBF)}{https://www.bmbf.de/}, grant number \grantnum{01IS20075}{01IS20075}.
A. Bulling was funded by the European Research Council (ERC; grant agreement 801708). 
The researchers from Augsburg University were partially funded by the German Ministry for Education and Research (BMBF), grant numbers \grantnum{16SV8688}{16SV8688} and \grantnum{16DHB2215}{16DHB2215}.
\end{acks}

\newpage
\bibliographystyle{ACM-Reference-Format}
\balance
\bibliography{bibliography}

\end{document}